\documentclass[12pt,preprint]{aastex}

\shorttitle{Dust Reddening in SDSS Quasars}
\shortauthors{Hopkins et al.}
\newcommand{\etal}{et al.}
\newcommand{\sigrat}{\frac{\sigma_{\rm Gaussian}}{\sigma_{dust}}}
\begin{document}
\title{Dust Reddening in Sloan Digital Sky Survey Quasars}
\author{Philip F. Hopkins\altaffilmark{1}, Michael A. Strauss\altaffilmark{1}, 
Patrick B. Hall\altaffilmark{1}, Gordon T. Richards\altaffilmark{1},
Ariana S. Cooper\altaffilmark{2},
Donald P. Schneider\altaffilmark{3}, 
Daniel E. Vanden Berk\altaffilmark{4}, 
Sebastian Jester\altaffilmark{5},
J. Brinkmann\altaffilmark{6}, and Gyula P. Szokoly\altaffilmark{7}
}
\altaffiltext{1}{Princeton University Observatory, Princeton, NJ 08544, USA}
\altaffiltext{2}{Joseph Henry Laboratories, Princeton University, Princeton, NJ
08544}
\altaffiltext{3}{Department of Astronomy and Astrophysics, 
Pennsylvania State University, University Park, PA 16802, USA}
\altaffiltext{4}{Department of Physics and Astronomy, University of Pittsburgh,
Pittsburgh PA 15260, USA}
\altaffiltext{5}{Fermi National Accelerator Laboratory, P.O. Box 500,
  Batavia, IL 60510} 
\altaffiltext{6}{Apache Point Observatory, P.O. Box 59, Sunspot, NM 88349, USA}
\altaffiltext{7}{Max-Planck-Institut f\"ur extraterrestrische Physik, 
Postfach 1312, 85741 Garching, Germany}

\begin{abstract}
We explore the form of extragalactic reddening toward quasars using a
sample of 9566 quasars with redshifts $0<z<2.2$, and accurate optical
colors from the Sloan Digital Sky Survey (SDSS). We confirm
that dust reddening is the primary
explanation for the red ``tail'' of the color distribution of SDSS
quasars. 
Our fitting to 5-band photometry normalized by the modal quasar color as
a function of redshift shows that this ``tail'' is well described by SMC-like 
reddening but not by LMC-like, Galactic, or Gaskell \etal\ (2004) reddening.
Extension
to longer wavelengths using a subset of 1886 SDSS-2MASS matches
confirms these results at high significance.  We carry out Monte-Carlo
simulations that match the observed distribution of quasar spectral
energy distributions using a Lorentzian dust reddening distribution; 
2\% of quasars selected by the main SDSS targeting algorithm (i.e.,
which are not extincted out of the sample) 
have $E_{B-V} > 0.1$; less than 1\% have $E_{B-V} > 0.2$, where the
extinction is relative to quasars with modal colors. 
Reddening is uncorrelated with the presence of
intervening narrow-line absorption systems, but reddened quasars are much more
likely to show narrow absorption at the redshift of the quasar than
are unreddened quasars.  Thus the reddening towards quasars is
dominated by SMC-like dust at the quasar redshift. 
\end{abstract}


\section{Introduction}

The discovery of a very red population of radio-selected 
quasars, and studies of the color distribution of 
quasars \citep{webst,broth,franc,richa,gregg,white} suggest that a
significant population of red quasars exists.
This red population could arise from several sources: an 
intrinsically red continuum, an 
excess of synchrotron emission in the red, intervening absorption by
galaxies along the line of sight, and dust extinction in the host
galaxy or central engine of the quasar itself. With its small
photometric errors and very large quasar sample selected in the
$i$-band, the Sloan Digital Sky Survey (SDSS; \citealt{york}) is
well-suited to explore the properties of mildly dust-reddened quasars.
\citet{richc} used redshift-independent criteria to quantify the
``redness'' of SDSS quasars based on their optical colors, and concluded
that the core of the color distribution was consistent with a range of
intrinsic slopes of quasar continua, while the extended red tail was
the sign of dust located predominantly at the redshift of the quasar.  
We extend this work in the
present paper, looking both for the change in slope and the induced
curvature expected in the quasar spectral energy distribution in the
presence of dust reddening.  Our aim is to 
characterize the nature of the dust giving rise to the observed
reddening, and to constrain its intrinsic distribution.
For example, \citet{gask} have recently proposed that quasars have a
nuclear reddening curve which is flat in the ultraviolet, in contrast to the
well-studied curves of the Milky Way (MW), the Large Magellanic Cloud (LMC),
and the Small Magellanic Cloud (SMC).

The SDSS and the sample
of quasars we use are described in \S~\ref{sec:sample}.  We fit the colors to
derive the slope and curvature in \S~\ref{sec:fits}, and compare our
results with simple models for the distribution of dust extinction.
Our conclusions are presented in \S~\ref{sec:conclusions}.

\section{The Sloan Digital Sky Survey and Sample Selection}
\label{sec:sample}

The quasars analyzed here were selected from the SDSS imaging survey,
which uses a wide-field multi-CCD camera \citep{gunn} which produces
accurate photometry in five broad bands ($ugriz$) which cover the
optical window from 3000 to 10,000\AA.  Quasar candidates are selected
\citep{richb} based on
their $ugriz$ colors \citep{fukug,luptb,hogg,smith,stoug} and
as optical matches to quasars from the VLA ``FIRST'' survey
\citep{becke}, and are spectroscopically observed in the wavelength range
$3800 <\lambda<9200$\,\AA. The quasar selection is not limited to UV-excess
objects; all objects lying sufficiently far from the stellar locus in
color-color space brighter than the $i$ band magnitude limit are selected,
with the exception of objects in certain exclusion boxes; see 
\cite{richa}.

Point-spread-function magnitudes were extracted as
explained by \citet{lupta}, and magnitudes have been
corrected for Galactic extinction using the maps of \citet{schle}. 
See \citet{stoug} for details of the spectroscopic
system and analysis pipelines. Astrometric calibration is detailed in
\citet{pier} and the spectroscopic plate tiling procedure in \citet{blanton}.

The sample consists of objects from the SDSS first data release 
(DR1; Abazajian \etal\ 2003) 
cataloged as quasars by 
\citet{schne}.
Objects with $z > 2.2$ are not included, in order to avoid contamination of the analysis of
spectral shapes and potential reddening by the effects of 
Ly$\alpha$ forest absorption.  We have included objects showing broad
absorption lines in their spectra, although we leave to future work to
explore whether their dust properties differ from those of ``normal''
quasars.  

At low redshift ($z \lesssim 0.3$), most SDSS AGNs have low luminosity
and do not strongly outshine their host galaxies.  Indeed, many of the
reddest objects are dominated by the spectrum of the host galaxy,
showing strong stellar absorption lines. To eliminate the potentially 
contaminating objects, we retained only objects targeted as 
low-$z$ quasars selected by the final version of the quasar target
selection algorithm (with $M_i < -22$ in a cosmology in
which $H_0 = 70$ km s$^{-1}$ Mpc$^{-1}$, $\Omega_M = 0.3$,
$\Omega_\Lambda = 0.7$, and in which quasar SEDs have
$f_{\nu}\propto\nu^{-0.5}$). Visual inspection confirmed that the
eliminated 
objects were primarily objects in which the host galaxy dominated the
spectrum, and that there are only a few objects in the retained sample
which show stellar absorption systems in their spectra.  The final 
sample consists of 9566 quasars. 

This sample was matched to the Two Micron All Sky Survey
(2MASS; \citealt{skrut})
all-sky point source catalog using a matching radius of 3$''$
\citep{schne}.  The addition of 2MASS observations extends the
wavelength baseline by adding $J$, $H$, and $K$ band photometry. 
Typical magnitude errors in each band are $\sim 0.1 - 0.15$ mag. This
sample again includes only objects with $0 \leq z \leq 2.2$, and the
final sample includes 1886 quasars.

\citet{richc} show that for $E_{B-V}=0.1$, the completeness of the
SDSS quasar target selection is at least 90$\%$ at $z<2$ for objects
luminous enough that extinction does not cause them to fall below the magnitude limit.
In this paper, we will be sensitive to objects with moderate amounts of dust
reddening, as objects with extreme dust reddenings ($E_{B-V}\gtrsim0.5$) will
be difficult to recognize as quasars and will usually be extincted out of the
sample.  The fraction of quasars which are so heavily reddened has been
estimated as 16\% based on SDSS data \citep{richc} and 15\%-22.5\%
based on a 2MASS+VLA-FIRST survey \citep{glikman04}.

\section{Fitting Reddening Models to Quasar Colors}
\label{sec:fits}

\subsection{Colors as a Function of Redshift}

We follow the convention of Richards \etal\ (2001; 2003), and define
``red'' quasars using colors relative to those of typical quasars as a
function of redshift. These papers used median colors
as a function of redshift; however, a population of dust-reddened
quasars will shift the median quasar colors redder, but will leave the
mode unaffected.  Thus we use the modal colors as a
function of redshift in this paper.  The DR1 quasars were sorted by redshift and then split into
bins of width 0.05 in redshift $z$. Each quasar's color was taken to
be a Gaussian centered at the given value for that quasar, with a
standard deviation equal to the photometric error for that particular
quasar and color. All Gaussians in the bin were then summed and the
peak of the resulting distribution was used as the modal color of the
bin. Bin sizes as large as 0.1 gave similar results, whereas bins 
with a significantly smaller numbers of quasars per bin gave noise-dominated results. 

The modes of the 2MASS colors $J-H$, $H-K$, and $J-K$ were found using
a sample of 5162 quasars from Table 1 of Veron-Cetty
\& Veron (2001) which have 2MASS-point source catalog counterparts within 4\arcsec.
A false coincidence rate of only 1.6\% is expected with this matching
radius \citep{bark}. 
Due to the smaller number of quasars,
the bin size was taken to be 0.1 in redshift. The modes in $z-J$ and
$g-K$ were taken 
from the SDSS-2MASS matched sample, and a bin of 0.2 was used due to
the still smaller sample size.
Table~\ref{modestbl} and Figure~\ref{modesfig} show the modal
colors as a function of redshift for the four SDSS colors ($u-g$, $g-r$,
$r-i$, $i-z$), three SDSS-2MASS matched colors ($z-J$, $J-H$,
$H-K$), and three colors commonly used as reddening indicators ($g-i$, 
$J-K$, and $g-K$, a color very close to $B-K$).
We show for comparison the median colors of Richards \etal\ (2001), which
tend to be biased to the red (by $\sim 0.025$ mag), as
expected for a distribution with a red tail.

\subsection{Fit Construction}

  We define {\em relative} colors of each quasar by subtracting the
modal color at that redshift from the quasar's observed colors; this
corrects the colors for the effects of emission lines redshifting
through the SDSS and 2MASS filters; in essence, this applies a
$k$-correction for the emission lines. 
 
Adding these relative colors to a fixed $i$ magnitude ($i=0$) gives
relative $ugrizJHK$ magnitudes: a quasar at the mode of the
color-redshift relation will thus have all relative magnitudes equal
to zero.  Deviations of the modal magnitudes from zero can be modeled
as due to reddening; as reddening is greater at shorter wavelengths,
we expect a curvature in the spectral energy distributions.
To quantify this, we fit a second-order Chebyshev polynomial to
the relative magnitudes of each quasar in turn as a function of the
logarithm of wavelength.  The effective wavelengths of the filters are
3541, 4653, 6147, 7461, 8904 {\AA}, respectively, in $ugriz$, and  12350, 16620, and
21590 {\AA}, respectively, in
$JHK$ (the variation in effective wavelength with the shape of the SED
is negligible for our purposes).  We fit the photometry rather than
the SDSS spectra to measure 
the reddening because the photometry has higher signal-to-noise ratio,
better calibration (cf., \citealt{DR3}), and a longer wavelength
baseline.

The Chebyshev fit to each quasar is carried out with $\chi^2$ minimization, 
using the PSF magnitude errors in each band (with a floor of 0.01
mag). 
The Chebyshev polynomials are designed to give
statistical independence to the derived slope and curvature for each quasar.  We use
the notation $c_2$, $c_1$, $c_0$, for the best-fit coefficients of the
Chebyshev polynomials of second, first, and zeroth order,
respectively.  Simple changes in the power-law slope will affect
$c_{1}$, while $c_{2}$ will remain close to zero. Dust reddening
causes curvature in the spectrum, and should give positive values of
$c_{2}$ correlated with $c_{1}$. A ``normal'' quasar with colors close
to the mode will have relative magnitudes all close to zero, and thus
$c_1 = c_2 = 0$.  In what follows, we will show results from our full
sample of 9566 quasars, where the fit is restricted just to the SDSS
filters, and 
to the subset with 2MASS photometry, where the fit includes $JHK$ as
well. 

The data do indeed call for a curvature ($c_{2}$) term. Figure~\ref{errfig}
shows the improvement in $\chi^2$ in second-order fits to the SDSS
photometry alone, relative to a linear fit.  The addition of a
single fitting parameter typically improves $\chi^2$ by of order
unity if the data are well-fit without this parameter.  For objects
with large $c_2$, the improvement in $\chi^2$ is typically much greater
than unity.  

\subsection{Fit Results and Comparison to Dust Reddening Models}

Figure~\ref{chebdr1} shows the
relationship between curvature $c_2$ and slope $c_{1}$ for the objects
in our full sample, fitting to the SDSS photometry alone.  Again,
the modal quasar has $c_1 = c_2 = 0$ at all redshifts.  The distribution of
the quasars is shown as contours and outlying points, with a
symmetric distribution about a curvature and slope of 0, and a tail.
Note the strong correlation between $c_1$ and $c_2$ in the tail, as
expected with dust reddening. 

We compare the observed correlation of $c_1$ and $c_2$ with those
derived from SMC-like, LMC-like, MW-like (Galactic), and \citet{gask}
dust reddening laws.  
We use the numerical formulations of \citet{pei} for the Galactic and
Magellanic laws. (There is, of course, considerable spatial 
variation in the dust extinction curve of the Milky Way and of the LMC 
\citep{draine}; the curves we refer to in this paper can be thought of as the
``traditional'' MW, LMC 30 Doradus region, and SMC ``bar'' extinction curves.)
For each reddening law, we simulate reddened quasar
colors by beginning with a `quasar' with all four SDSS modal colors set to 0.
We then determine the changes to the colors that would result from applying
each of the four reddening laws with $0\leq E_{B-V}\leq0.8$ (in bins of 0.05) to
the modal quasar at $0 \leq z \leq 2.2$ (in bins of 0.2).  In all there are 204
such simulated colors for each reddening law (17 $E_{B-V}$ bins times 12
redshift bins). The reddening we apply here is
relative to the modal quasar, and therefore is in addition to any
reddening experienced by the modal quasar itself.
We explicitly modeled extinction only at the quasar redshifts
(cf., \S~\ref{sec:abslines} below),  However, the relative colors we use are
independent of redshift, so a quasar with associated dust at $z=1$ and
a quasar at $z=2$ with the same amount of intervening dust at $z=1$
both have the same shift in the $c_1$ vs. $c_2$ diagram.  That is, our
simulated $c_1$ and $c_2$ values describe the effects of both
associated and intervening dust.  

These simulated relative colors were then fit to the Chebyshev
polynomials to obtain values of the slope and curvature ($c_1$ and
$c_2$).
The results are shown in Figure~\ref{chebdr1}. Model quasars
reddened by an SMC-like law appear as solid 
dots, those reddened by an LMC-like law as 
crosses, those reddened by a MW law as 
triangles, and those reddened by the Gaskell \etal\ (2004) law as plus
signs.  The SMC law is nearly a power law in
wavelength, and in particular does not have a 2175\AA\ feature, which
is why all the model points lie along a single locus independent
of redshift.  

The observed distribution is extended along the axis predicted by an
SMC-like dust reddening law.  Although objects reddened by LMC-like, 
MW-like, or \citet{gask} laws can appear in the tail, the majority of
objects with these reddening laws fall far from the observed tail.  For additional emphasis,
the simple least-squares linear fit to the simulated quasars reddened by
the SMC-like law is shown in the $c_{1}$, $c_{2}$ space.  Although our
data do not exclude the possibility that there is some component of
dust reddening in at least some of our objects which follows other
reddening laws, it appears that SMC-like
extinction curves dominate the reddening along most quasar
sightlines.  Figure~\ref{fig:extinction_fit} makes explicit the relationship
between reddening, $E(B-V)$, and $c_1$ and $c_2$ for the SMC law;
points of a given $E(B-V)$ at different redshifts are connected.  The
equivalent plot for the MW extinction law is not at all linear,
because of the influence of the 2175\AA\ feature. 

This analysis has focused purely on the {\em reddening} of
quasars; we have not modeled here the {\em extinction} of quasars
which would take them below our magnitude limit, causing them to drop
out of the sample. However, extinction does not affect our conclusion
about the dominance of SMC-like reddening along quasar sightlines.
At $z<2.2$, the $i$-band selection of the SDSS corresponds to selection
at rest-frame $\lambda > 2330$\,\AA.  The SMC, LMC and MW extinction curves
for a fixed reddening $E_{B-V}$ agree within 10\% at such wavelengths.
Since the SDSS can detect intrinsically very luminous quasars with SMC-like
reddenings of up to $E_{B-V} \simeq 0.5$ \citep{hallb}, it can also detect
quasars with similar reddenings but LMC-like or MW-like extinction curves.
Thus the absence of highly reddened quasars with the slope and curvature parameters
expected for LMC-like or MW-like extinction curves must be due to the scarcity
of dust with such extinction curves along the sightlines to
quasars.

The column density of gas $N_H$ required to generate a given
dust reddening $E_{B-V}$ {\em is} a function of the extinction curve.
Because the SMC-LMC-MW extinction law sequence is 
probably at least in part a metallicity sequence, 
MW-like dust has $E_{B-V}/N_H$ approximately
nine times larger than SMC-like dust; and LMC-like dust has $E_{B-V}/N_H$ 
approximately 2.25
times larger than SMC-like dust \citep{diplas,koorn,bouchet}.
Thus, the limiting $E_{B-V}\simeq0.5$ to which the SDSS is sensitive 
will be reached at a lower column density for gas with MW- or LMC-like 
extinction curves than for SMC-like extinction curves. Since lower column density
systems are likely to be more common than higher column density ones, the slope-curvature
plane would show an excess of MW- and LMC-reddened objects even if all three
reddening curves were equally common. The relative lack of such objects in spite
of this effect means that SMC-like extinction is truly dominant along quasar sightlines.

The Chebyshev fitting procedure was next applied to SDSS-2MASS matched quasars. Using the
SDSS photometry for $ugriz$ magnitudes, and the 2MASS photometry for
$JHK$ magnitudes, each of the 1886 quasars in this sample had seven colors
instead of the four for SDSS-only quasars (adding $z-J$, $J-H$, and
$H-K$). The baselines are extended by a factor of $\sim 3$ at the cost of
larger photometric uncertainties, but this increases 
the signal-to-noise ratio of the Chebyshev
coefficients by a factor of $\sim3$.  SDSS
colors were normalized using the modes derived from the DR1 quasar
sample, the $J-H$ and $H-K$ 2MASS colors were normalized using the 
modes derived from the Veron-Cetty
\& Veron (2001) 2MASS quasar sample, 
and $z-J$ colors were normalized using modes
derived only from the 1886 quasar SDSS-2MASS matched sample
(Figure~\ref{modesfig}).  
The Chebyshev polynomials were recalculated for the new 
baselines and then used as above to fit to the relative colors. The
results are plotted in Figure~\ref{cheb2ma}. Again, ``modal'' quasars
with $0 \leq z \leq 2.2$ and $0 \leq E_{B-V}\leq 0.8$ were simulated
using SMC-like, LMC-like, MW-like, and \citet{gask} dust reddening laws, and the
results are overlaid. The extension of the contours of the observed
distribution along the axis predicted by SMC-like reddening is again
strong, although the distinction between the reddening laws
is not as marked over this longer wavelength range (the strong curvature 
induced by the 2175 \AA{} ``bump'' is ``averaged out'' over the 
longer wavelength baseline).  

\subsection{Monte Carlo Simulation of the Reddening Distribution}

The $E_{B-V}$ values derived from direct fits to the photometry of
individual objects are degenerate with other parameters, such as
varying emission line strengths and the continuum power-law slope of
the quasar spectrum. Thus we cannot infer the distribution of
$E_{B-V}$ directly from these fits.  Instead, we simulate the quasar population
with various models for reddening and compare with the observed
population.  In this analysis, we are just modeling the effects of
reddening; we do not attempt to include the effects of extinction,
which would cause objects to drop out of the sample.

For each quasar in the sample, we take its redshift as given, and
start by setting its colors to the mode at that redshift.  We model
three contributions to the relative colors: deviations in the
continuum power-law slope of the intrinsic spectrum, SMC-like dust
reddening at the quasar redshift, and photometric deviations from 
the ``modal'' quasar. For each quasar, we generate a
random intrinsic slope deviation, $E_{B-V}$ value, and photometric
deviations, each according to an assumed probability distribution. We
assume that these probability distributions are independent of
redshift and one another.  

The photometric deviations model includes uncertainties in the
determination of the modal colors and intrinsic deviations about those
modal colors (e.g.\ from variations in emission line strength), as well as the
PSF magnitude errors for each object.  We assume the photometric
deviations to be normally distributed around zero, with standard deviation
($\sigma_{error}$), to be determined.  We follow \citet{richc} and
assume that the deviations from the modal power law slope $\alpha$ are
also normally distributed around zero, with standard deviation
$\sigma_{\alpha}$ as a quantity to be determined.

The distribution of $E_{B-V}$ values is also peaked at zero, since the
``modal'' quasar always has relative colors equal to zero, and as such
shows no curvature as is induced by dust reddening. The existence
of the ``red tail'' in the quasar color distribution suggests that the
distribution must be asymmetric about zero.  Given the distribution of
fitted $E_{B-V}$ values, degeneracies between $E_{B-V}$ and power-law
slope at small values of $E_{B-V}$, and the existence of some
significantly bluer-than-average quasars, we allow negative $E_{B-V}$
values, which correspond to quasars which have undergone less dust
reddening than the modal quasar at their redshift. We assume that the
negative half of the $E_{B-V}$ distribution is also Gaussian with
standard deviation $\sigma_{\rm Gaussian}$, giving a rapid falloff at large
negative values as we expect. The high-curvature tail is to be
explained by dust reddening, so the falloff
with positive $E_{B-V}$ must be less rapid than Gaussian. We test two
forms for $P(E_{B-V}>0)$:
\begin{eqnarray} 
P(E_{B-V}>0) \propto \exp(-x)\\
P(E_{B-V}>0) \propto \frac{1}{1+x^{n}}
\end{eqnarray}
where $x \equiv \frac{E_{B-V}}{\sigma_{dust}}$
and $n$ and $\sigma_{dust}$ are treated as quantities to be
determined. As this model is purely empirical, there is no reason that 
the characteristic width of the Gaussian ($E_{B-V}<0$) half of the
distribution should be the same as that of the $E_{B-V}>0$
half, so we also fit for their ratio, $\sigrat$.
Fitting to the histogram of $\Delta(u-g)$ gives an approximate
Lorentzian ($n = 2$) profile, with $\sigma_{dust}\sim 0.02$, so we
take these values as initial guesses.

Once the entire model sample has been fitted, a two-dimensional
Kolmogorov-Smirnov (K-S) 
test \citep{fasano,peacock} was used to compare the results to the
original distribution.  The $D$-value is the 
largest fractional difference in the population of any one quadrant,
after dividing the slope-curvature space into four quadrants at each
point in turn and comparing the results; it quantifies the goodness of
fit.

An iterative process was used to obtain final values for $n$,
$\sigma_{dust}$, $\sigma_{error}$, $\sigrat$, and
$\sigma_{\alpha}$. Several values of each parameter were given,
each with fractional deviation $\sim 0.5$ from the initial values
mentioned above, and the full simulation was repeated for all possible
combinations of these values. The combination of values which gave the
lowest $D$-value was taken as the seed for the next iteration. The same
number of possible values for each parameter was tested with each
subsequent iteration, but the range of the parameters (fractional
deviation from the best value of the previous run) was decreased by a
factor of 2 and was re-centered on the minimum $D$-value parameter
combination of the previous iteration.  The iteration continued until
the fractional change in the minimum $D$-value of subsequent iterations
was less than $10^{-3}$. The entire process was repeated with
identical and different initial values to check the stability of the
convergence.

The results of these simulations are shown in Figures~\ref{simdr1}
and~\ref{sim2ma}.  The figures show the contours of the observed
distribution and those of the simulated distribution using the final
values of $n$, $\sigma_{dust}$, $\sigrat$, 
$\sigma_{error}$ and $\sigma_{\alpha}$. The
best-fit simulation average value of the K-S statistic $D$ is shown in
Table~\ref{simtbl}, along with the best-fit values of the simulation
parameters for each sample. We find for the SDSS sample a marginally
better fit using the exponential profile, with 
$\sigma_{\alpha}\approx0.11,\,\sigma_{error}\approx0.065,
\,\sigma_{dust}\approx0.032\,{\rm and}\,\sigrat\approx0.54$.
The errors given in Table~\ref{simtbl} represent the range of
tested values which give $D$-values within $\pm 0.01$ with all other
parameters held constant; these errors should be considered as
heuristic only. We find similar values for the matched SDSS-2MASS 
sample, except for the much larger $\sigma_{error}\approx0.18$
due to the larger 2MASS photometric errors. 
In both cases, the best-fit pseudo-Lorentzian profile is similar to 
the best-fit exponential profile. 

The values of $\sigma_{\alpha}$ and $\sigma_{error}$ are
reasonable.  \citet{richc} find 
$\sigma_{\alpha}=0.125$, between
the values obtained for the SDSS and SDSS-2MASS samples and within 
the error bounds obtained.
The quantity $\sigma_{error}$ is only a factor of two above the typical
errors in the PSF magnitudes, due to the effects of photometric
calibration uncertainty, possible errors in the determination of
the modal colors with redshift, and intrinsic deviations about the mode 
from sources other than variations in slope and dust reddening.

The best-fit model is $\sigma_{dust} = 0.03$ with $\sigrat=0.55$; the
parameters are essentially identical for the two parameterizations of
Equations 1 and 2.  
In this model, we expect less than 1\% of all quasars in the sample to have $E_{B-V}>0.2$, 
2\% to have $E_{B-V}>0.1$, and $9_{-2}^{+4}$\% to have $E_{B-V}>0.055$.  
Richards \etal\ (2003) estimated that $\sim 6$\%
of quasars at the observed SDSS flux limit have
$E_{B-V}>0.055$, but given the large errors, this value is consistent
with the above estimates. The structure of the profile is such that the fraction of quasars with 
$E_{B-V}>0.055$ is highly sensitive to the exact value of $\sigma_{dust}$.
As Figures~\ref{simdr1} and \ref{sim2ma} show, this model does an excellent
job of fitting the tails of the marginal distributions in $c_1$ and $c_2$.
A quasar at $z = 1$ with $E_{B-V}$ of 0.2 has an extinction 
in the $i$ band of 0.95 magnitudes for SMC-like reddening. 
Thus, the SDSS probes one magnitude
fainter for unreddened quasars than for quasars reddened by $E_{B-V}=0.2$.
Quasars with reddening as large as $E_{B-V} = 0.5$ are very rare in our data,
mostly because the surface density of quasars luminous enough to be reddened
so heavily and still lie above the SDSS magnitude limit is very small.
 
\subsection{Where is the Absorbing Material?}
\label{sec:abslines}

The characteristic extinction we find along quasar sightlines,
$\sigma_{dust}=0.03$, corresponds to a characteristic column density
of $N_H=1.31\pm0.33 \times 10^{21}$\,cm$^{-2}$ for SMC-like extinction
\citep{bouchet}, a level detectable in X-ray observations only for very nearby quasars or very
long exposures.  This value is at least an order of magnitude larger
than expected from intervening absorption alone, as follows.  If we make
the extreme assumption that the column density distribution of intervening
absorbers follows a power law of index $-1.2$ \citep{rao} up to the
SDSS reddened-quasar detection limit of $2.2\times10^{22}$\,cm$^{-2}$
(corresponding to $E_{B-V}=0.5$ for SMC extinction), then we find an
average total intervening column density of $N_H=4.7 \times 
10^{19}$\,cm$^{-2}$.  Therefore, reddening along quasar sightlines
is dominated by reddening at each quasar's redshift, whether in the
region of the central engine or elsewhere in the quasar host galaxy.
There are certainly examples of individual quasars
that are dominated by intervening reddening (e.g., Wang \etal\ 2004),
but they are exceptions to the rule.

We can test the conclusion that the reddening occurs in the host
galaxy of the quasar directly: 
the dust which causes reddening is often accompanied by gas which
gives rise to narrow interstellar absorption lines, such as the
MgII and CIV doublets.  Quasar spectra
often show such narrow lines, both at the redshift of the quasar (and
therefore presumably due to the host galaxy of the quasar) and at
intervening redshifts (cf., the SDSS study by Nestor \etal\ 2003).  We
therefore examined the spectra of the 393 quasars in the sample with
$c_1 < -2$ (the reddened sample).  For each of the quasars in the
reddened sample, we found an unreddened quasar (i.e., with $0 < c_1 <
0.5$) which closely matched in redshift, thus generating the control
sample.  

In the redshift range over which Ca H and K were visible ($z
< 0.7$), we found no difference in the presence of these lines,
indicating that stellar continuum contamination was not the cause of
the reddening.
For redshifts $z > 0.4$, the MgII doublet becomes visible in the SDSS
spectra; at $z > 1.5$, the CIV doublet appears as well.  The
SDSS spectral resolution $\lambda/\Delta \lambda = 2000$ is more than
adequate to resolve both these doublets in absorption, so they are
easy to recognize.  305 of the
quasars in the sample fell within this redshift range.  Of these, 127
showed intervening MgII or CIV absorption in the reddened sample,
statistically identical to the 125 objects in the control sample with
intervening metal lines.  Therefore, we saw no correlation between the
presence of reddening and the presence of intervening absorption
systems (see Murphy \& Liske 2004 for a similar result).  However, 56
of the reddened quasars showed metal-line 
absorption at the redshift of the quasar itself, while only 12 of the
objects in the control sample showed self-absorption.  Thus there is
strong empirical evidence that the reddening we observe is due to 
dust at the redshift of the quasar.

\subsection{Composite Spectra Construction}

\citet{richc} discuss composite quasar spectra in bins of the relative
$g - i$ color, and thus as a function of slope ($c_{1}$). Here, we
create six composite spectra in bins of distance along the SMC-dust
reddening axis $d_{SMC}$ in the $(c_1,c_2)$ diagram. The composites are
constructed in the same manner as the \citet{vande} SDSS quasar
composite, using a similar code. The quasars are shifted to their rest
frame wavelengths, rebinned to a common wavelength scale, scaled by
the overlap of the preceding average spectrum, and weighted by the
inverse of the variance.  Extremely low signal-to-noise ratio points
and those masked as unreliable are discarded. The geometric mean of
the spectra was used, which preserves input power-law slopes
(and $E_{B-V}$ values, if all dust is at the quasar redshifts 
and the curvature in the reddening law is the same in all objects; 
see the Appendix to Reichard \etal\ 2003).

For the purposes of creating composite spectra with the maximum
possible signal-to-noise ratio, we extend our sample to a much larger
(but not yet finalized) sample of 44619 SDSS quasars, obtained using
identical selection criteria to the DR1 sample. First,
a sample spectrum was created for quasars close to the mode, from the
9626 quasars with $-0.1<d_{SMC}<0.1$. 
Five more composite spectra are created from the populations in bins of
increasing d$_{SMC}$, as shown in Table~\ref{spectbl}.  Quasars were
randomly removed from each bin to ensure that the distribution of
quasars with redshift was the same in each sample; the table gives
the resulting number of objects.

The resulting spectra are shown in Figure~\ref{specfig}. They have
been normalized to constant flux at 6000 {\AA}. The results are as
expected for dust reddening processes: the decrease in blue flux with
increasing distance along the axis of SMC reddening is very strong,
while the increase in red flux (due to anchoring the spectra at 6000
\AA) is comparatively weak.  
The upper panel
of this figure shows the ratio of the composite spectrum
to the modal composite spectrum reddened with the SMC reddening law
with the given $E_{B-V}$ value, determined by minimizing the $\chi^{2}$
of the absolute difference between the spectra.
We exclude emission line regions 
by using only the continuum windows 
$2100<\lambda<2600, 3000<\lambda<4200,$ and $5200<\lambda<6300$
\,\AA{} for the $\chi^{2}$ minimization.
  
The
ratio curves after correcting for the putative reddening are not
perfectly flat, especially in the narrow emission lines 
(perhaps indicating that the dust is distributed on scales smaller
than the narrow-line region, and is not pervasive through the host
galaxy).  There is also a systematic broad deviation from unity
centered around 3900\AA\ which remains unexplained.  But overall, these composite spectra 
 are well approximated by the modal composite
reddened by an SMC-like dust reddening law, and indicate that the
distance along the SMC-like reddening axis in the $c_1-c_2$ diagram is
a monotonic function of the degree of reddening in the spectra.  
Moreover, the composite spectra show no evidence of the 2175\,\AA{}
bump seen 
strongly in MW-like reddening laws and to a lesser extent in LMC-like
reddening laws. However, the 2175\,\AA{} bump due to
absorption from intervening systems has 
been seen in the spectra of a few individual objects 
\citep[][and references therein]{wang04}.

\section{Summary and Discussion}
\label{sec:conclusions}

We investigate the reddening law towards 9566 SDSS quasars, including
a subset of 1886 quasars matched to 2MASS by exploring the shapes of
the spectral energy distributions from broad-band photometry. To
remove the color changes with redshift induced by emission lines, we
subtract the modal colors of quasars as a function of redshift. We fit
a quadratic 
Chebyshev polynomial to the relative magnitudes as a function of
wavelength to derive the slope and curvature (\S~\ref{sec:fits}).  The
vast majority of quasars have a 
slope and curvature close to zero, i.e., an SED very similar to that of the
modal quasar. However, the slope-curvature distribution has an
extended tail in the direction of negative slope and positive
curvature, which is the sign of a population of quasars with
significant dust reddening.  We compare models of SMC-like, LMC-like, 
and MW-like dust reddening laws from \citet{pei} and
the \citet{gask} reddening law 
with the data, and find that only the SMC
model fits the data well. 

We carry out Monte Carlo simulations of the two-dimensional
distribution of slope and curvature, and show that the observed
distribution can be well fit with contributions to relative quasar
colors from photometric and modal errors, deviations in power-law
continua, and SMC-like dust-reddening. The distribution of $E_{B-V}$
values responsible for dust reddening can be well approximated with a
Gaussian-Lorentzian or Gaussian-exponential combination with standard
deviation $\sigma_{dust}= 0.03$. We again note that these values are
relative to the possible modal $E_{B-V}$ values of the quasar
population.  This dust-reddening is responsible for the non-symmetric
tail of the $c_{1}$-$c_{2}$ distribution. The fitted values for the
dispersion in the intrinsic slope of quasars, 
$\sigma_{\alpha }$ and the photometric error dispersion,
$\sigma_{error}$, dominate the slope and 
curvature dispersions, respectively, in the core of the relative color
distribution, and $\sigma_{dust}$ dominates the relative color
distribution of the tail.  Differential extinction between different
reddening laws is not a significant selection effect in our $i$-band
selected, $z<2.2$ sample.

Extension to larger wavelength baselines with a matched SDSS-2MASS
sample is consistent with the conclusion that SMC-like dust reddening is the
dominant cause of the reddening of the quasar population along our
lines of sight, with LMC-like dust reddening much less frequent and
MW-like or \citet{gask} reddening rare.  
The typical column density responsible for the reddening is at least an
order of magnitude larger than expected for intervening absorption
alone, and we have seen a strong correlation between dust reddening
and the presence of narrow absorption lines at the redshift of the quasar;
therefore, most of the extinction along quasar sightlines occurs in the
immediate nuclear environs or host galaxy of the quasar.  Dust
associated with the nucleus itself seems
more likely, however, given that quasar composite spectra show different
broad-line properties as a function of continuum color \citep{richc},
and that the strength of the narrow-line region seems to be
uncorrelated with reddening (Figure~\ref{specfig}).

We do not confirm the finding of \citet{gask} that the reddening law toward
quasars is flat shortward of $\sim$3750\,\AA{}.  Their reddening law was
constructed primarily using composite spectra of 72 radio-selected quasars,
supplemented by a composite of several hundred radio-quiet quasars.
We cannot rule out a small population of objects, radio-loud or otherwise,
reddened by dust with such an extinction law, but it is not the dominant
extinction law towards quasars.  It may be that \citet{gask} are interpreting
as reddening the difference between dominant, beamed, power-law continua in
radio-loud quasar spectra and the steepening of the continuum at
$\gtrsim$4000\,\AA\ observed in composite quasar spectra from several surveys
not dominated by radio-loud quasars
(see the discussion in \S\,5 of \citealt{vande}).

If we are correct that the bulk of the observed reddening towards quasars
occurs in their nuclear environments, rather than farther out in their host
galaxies, why should SMC-like dust extinction curves be prevalent when 
quasar nuclei are known to be typically quite metal-rich \citep{hf99}?  In
a recent study which concluded that SMC extinction is a good representation
of the optical and X-ray absorption towards a small sample of X-ray selected
quasars, \cite{will} suggested that this might be a coincidence.  It
may not be a coincidence, however, if the radiation 
field is important in determining the extinction properties of the dust.
Models of the SMC dust grain size distribution have fewer large silicate grains
and fewer small carbonaceous grains (PAHs) relative to the MW distribution.
The relative excess of small silicate grains produces a steeper UV extinction
curve, while the relative lack of PAHs of radius $\sim$30-150\,\AA\ explains
the weakness of the 2175\,\AA\ feature \citep{wd01}.
Such size distributions are consistent with theoretical predictions 
for dust exposed to high fluxes of ionizing radiation \citep{perna}.
However, those simulations predict a much lower ratio of silicate to
carbonaceous grains than required to reproduce an SMC-like extinction curve;
in fact, the simulated UV extinction curves are flatter,
not steeper, in harsh radiation environments.  
It remains to be seen what sort of self-consistent dust models can
reproduce our observations, but models incorporating the supersolar
abundances appropriate to quasars are an obvious first step.

We have not distinguished between normal quasars and
those with broad absorption lines.  It is well-known that BAL quasars
tend to be more strongly dust-reddened (e.g., Reichard \etal\ 2003 and
references therein).  We plan to explore this question in the future
with an expanded sample of the roughly 50,000 quasars with SDSS
spectra available to date. 

The analysis here has concentrated purely on the observed colors of
quasars; we have not attempted to model the effects of extinction
which would cause objects to fall out of our sample altogether.
In particular, this analysis yields no direct constraint on a population of
highly extincted quasars, as suggested,
e.g., by studies of radio-selected quasars \citep{white}, deep hard X-ray
surveys \citep{barger}, and Type II quasars \citep{zakamska}.  To
include the effects of extinction (both at the quasar redshift and 
from foreground systems) in this analysis will require a detailed analysis of
the luminosity function and the redshift evolution of the SDSS
quasar sample.

Funding for the creation and distribution of the SDSS Archive has been
provided by the Alfred P.  Sloan Foundation, the Participating
Institutions, the National Aeronautics and Space Administration, the
National Science Foundation, the U.S. Department of Energy, the
Japanese Monbukagakusho, and the Max Planck Society. The SDSS Web site
is http://www.sdss.org/.  The SDSS is managed by the Astrophysical
Research Consortium (ARC) for the Participating Institutions. The
Participating Institutions are The University of Chicago, Fermilab,
the Institute for Advanced Study, the Japan Participation Group, The
Johns Hopkins University, Los Alamos National Laboratory, the
Max-Planck-Institute for Astronomy (MPIA), the Max-Planck-Institute
for Astrophysics (MPA), New Mexico State University, University of
Pittsburgh, Princeton University, the United States Naval Observatory,
and the University of Washington.

MAS and PBH acknowledge support from NSF grants AST-0071091 and
AST-0307409.  We thank an anonymous referee for comments that
substantially improved the presentation of the paper.

\begin{center}
\begin{figure}
    \plotone{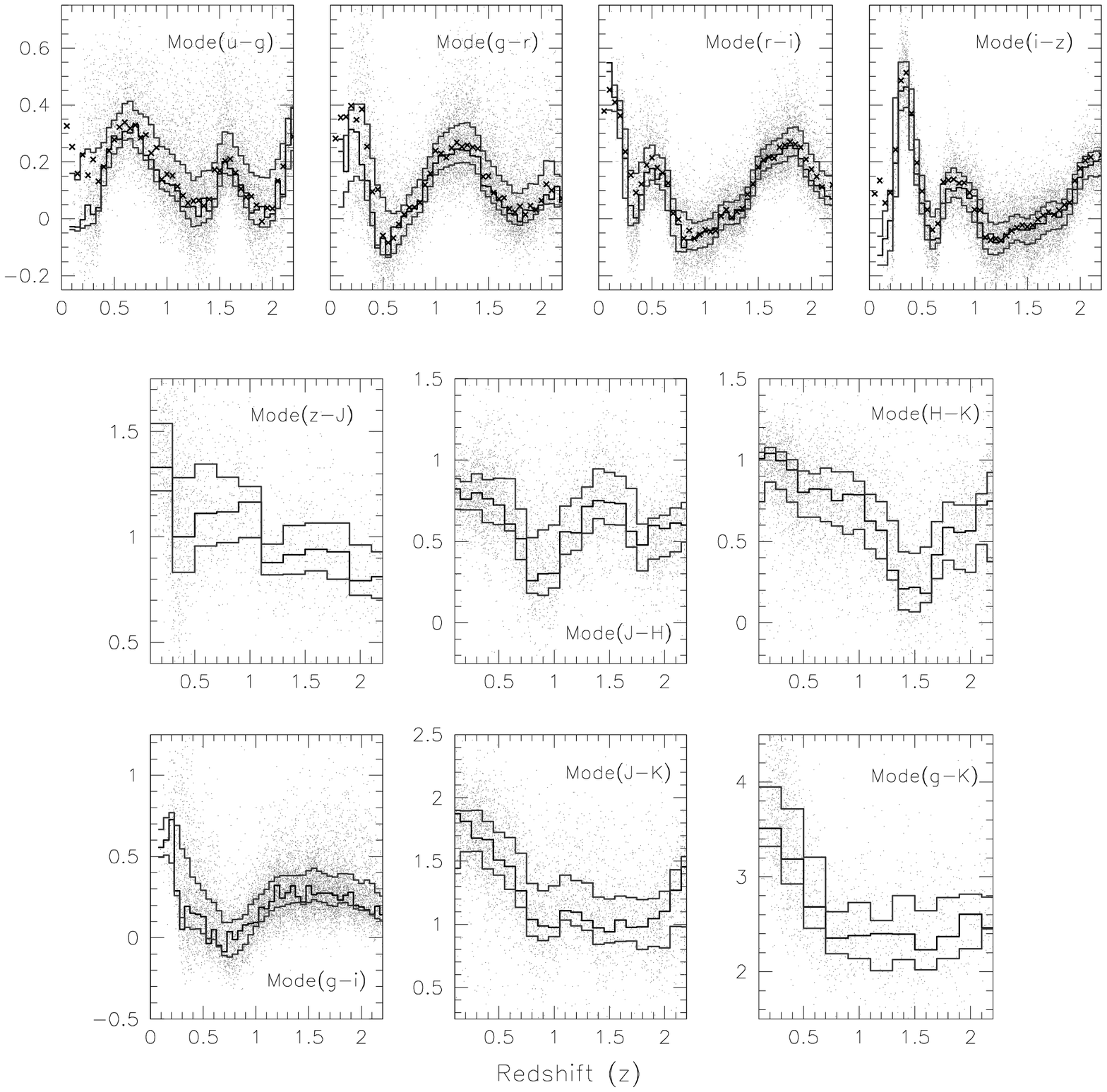}
    \caption{The dependence of modal colors on redshift. 
    Points plot individual quasar colors, and the lines in each box show the 
    modal color and the outer quartiles as a function of redshift. The
    top four boxes are the SDSS colors and the center three boxes are
colors obtained from the SDSS-2MASS 
    matched sample. The bottom three boxes show the commonly used reddening indicators 
    $g-i$, $J-K$, and $g-K$ as determined directly.
    The modes $u-g$, $g-r$, $r-i$, $i-z$, and $g-i$ were taken in bins of $\Delta z=\pm0.05$. Modes
    $J-H$, $H-K$, and $J-K$ use bins of $\Delta z=\pm0.1$, and $z-J$ and $g-K$ use bins
    of $\Delta z=\pm0.2$.
    For comparison, the median color-redshift relation of Richards \etal\ (2003) is shown 
    for the SDSS colors as $\times$'s.  
\label{modesfig}}
\end{figure}
\clearpage
\begin{figure}
    \plotone{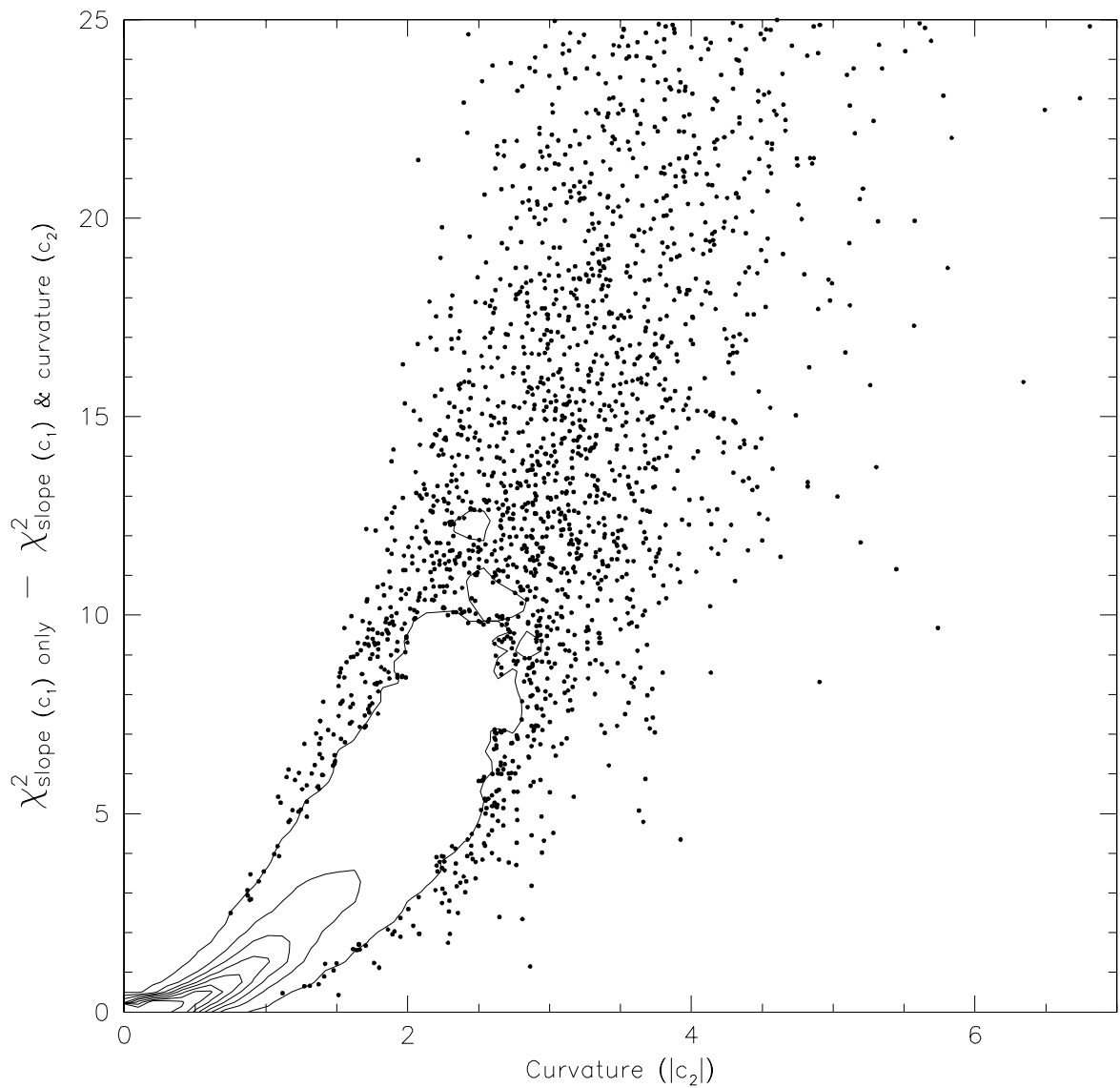}
    \caption{The magnitude of the curvature $c_2$ is plotted against the 
    improvement in $\chi^2$ upon the 
    addition of the second-order term to the fit to the SDSS photometry, for the 
    9566 DR1 quasars. Contours indicate the density of points. 
    Improvements greater
    than unity are statistically significant; for most of the
    high-$c_2$ objects, the addition of curvature to the model
    significantly improves the fit.\label{errfig}} 
\end{figure}
\clearpage
\begin{figure}
    \plotone{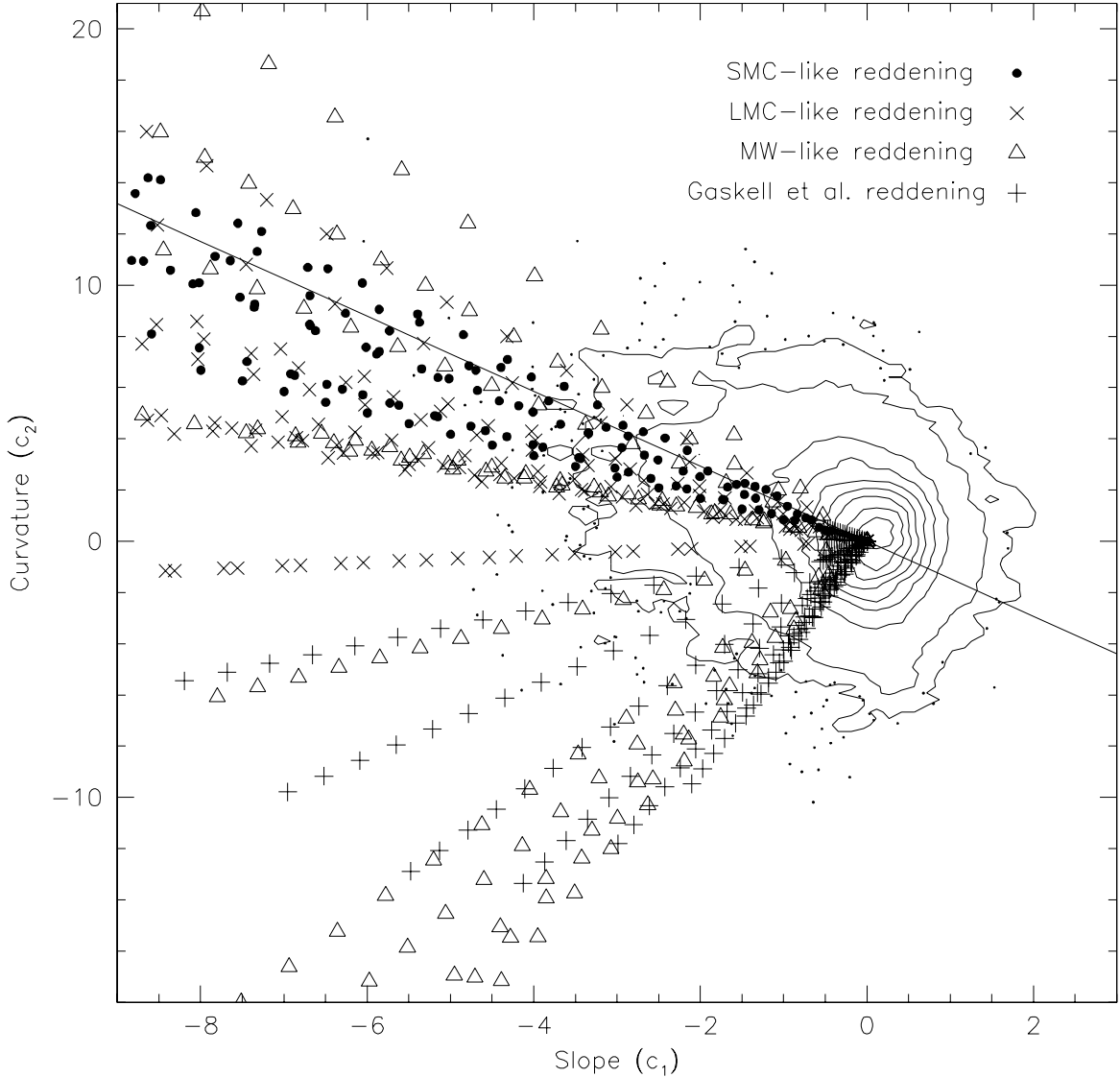}
    \caption{Curvature vs.\ slope ($c_2$ vs.\ $c_1$) for the 9566 DR1
      quasars; SDSS photometry alone is used. The distribution of the
      quasars is shown in the  
      contour map, with a symmetric distribution about the origin,
      except for a tail, as expected. Larger absolute values of each
      correspond to larger deviations from the flat median, and the
      tail in the negative slope, positive curvature direction
      corresponds to objects that are redder than average. Simulated
      quasars reddened by an SMC-like dust reddening law are shown as 
      dots (with the solid 
      line as the linear best-fit, including points beyond the edge of
      the plot), by an LMC-like law as 
      $\times$'s, by a MW-like law as 
      triangles, and by the \citet{gask} law as $+$'s.
      Points are simulated with $0\leq z\leq2.2$ and 
      $0\leq E_{B-V}\leq0.8$. The large range in LMC-like and MW-like points is 
      caused by the 2175\,\AA{} bump moving through the filters with redshift.
     \label{chebdr1}}
\end{figure}
\clearpage
\begin{figure}
\plotone{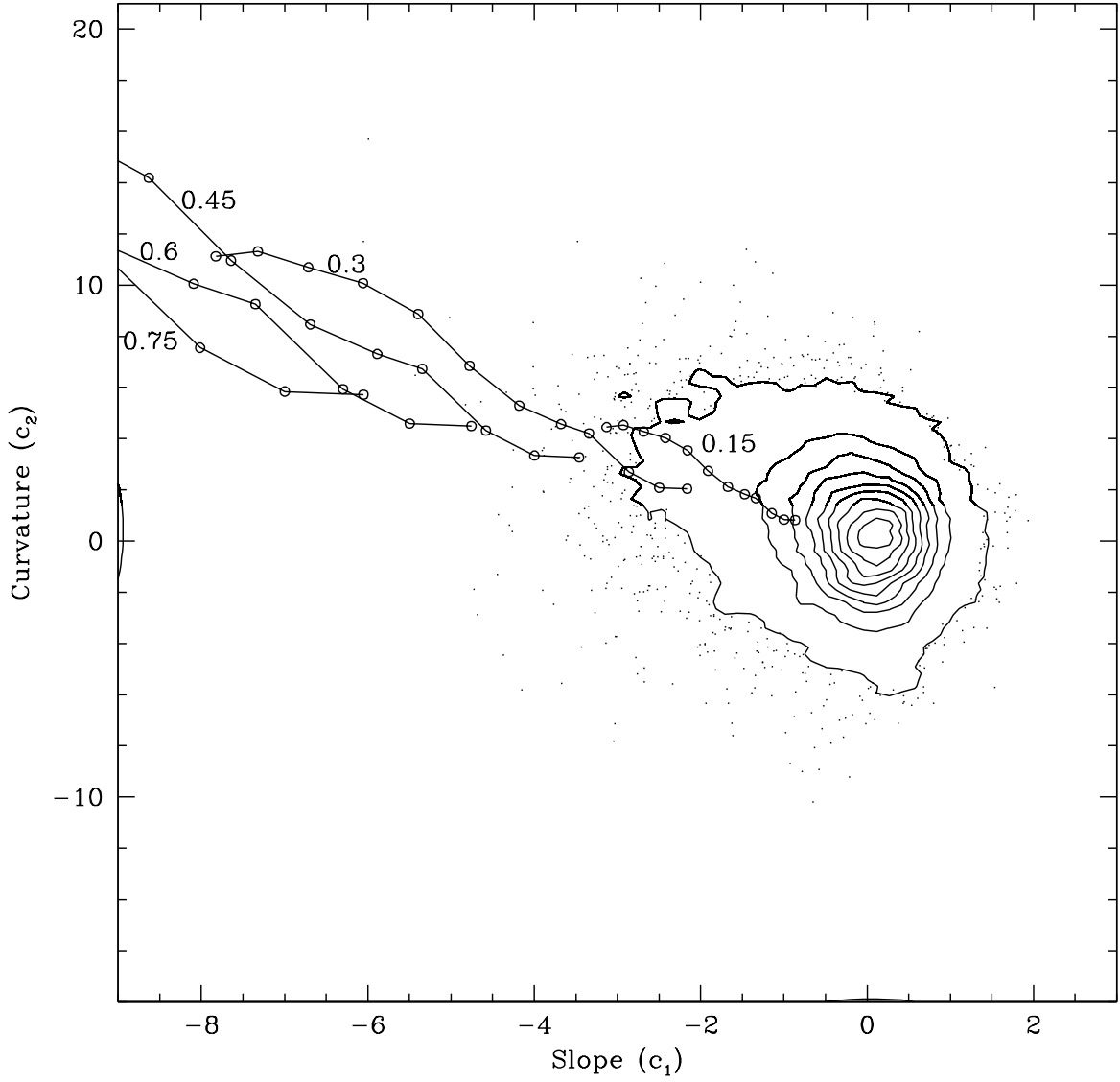}
\caption{As in Figure~\ref{chebdr1}, connecting the points of constant
  $E(B-V)$ for the SMC model points alone.  The curves are labelled
  with the value of $E(B-V)$, and an open circle is given every 0.2 in
  redshift for $ 0 < z < 2.2$.\label{fig:extinction_fit}}
\end{figure}
\clearpage
\begin{figure}
    \plotone{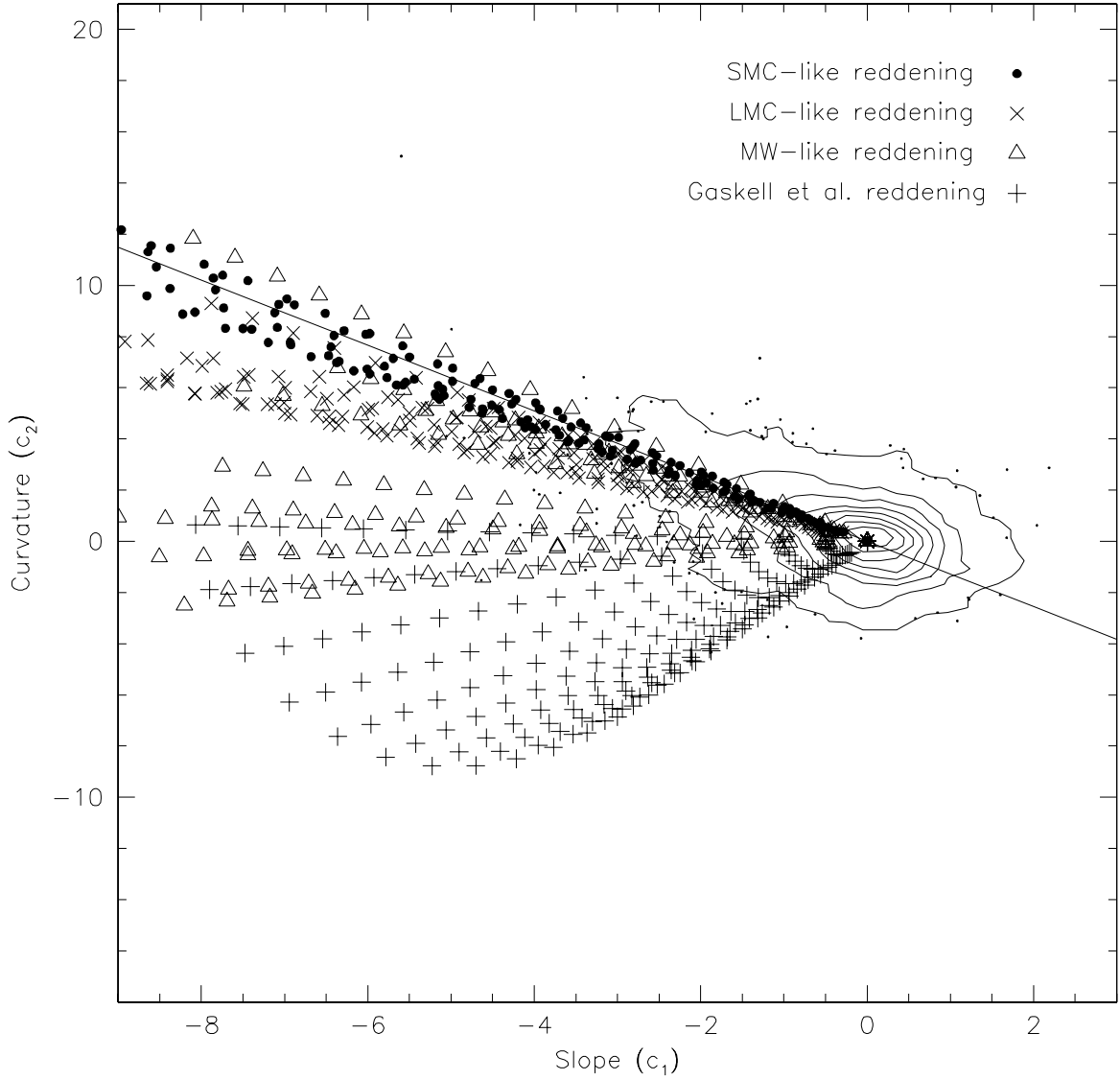}
    \caption{Curvature vs.\ slope ($c_{2}$ vs.\ $c_{1}$) for the 
    1886 SDSS-2MASS matched quasars. The format 
    is the same as that of Figure~\ref{chebdr1}.\label{cheb2ma}}
\end{figure}
\clearpage
\begin{figure}
    \plotone{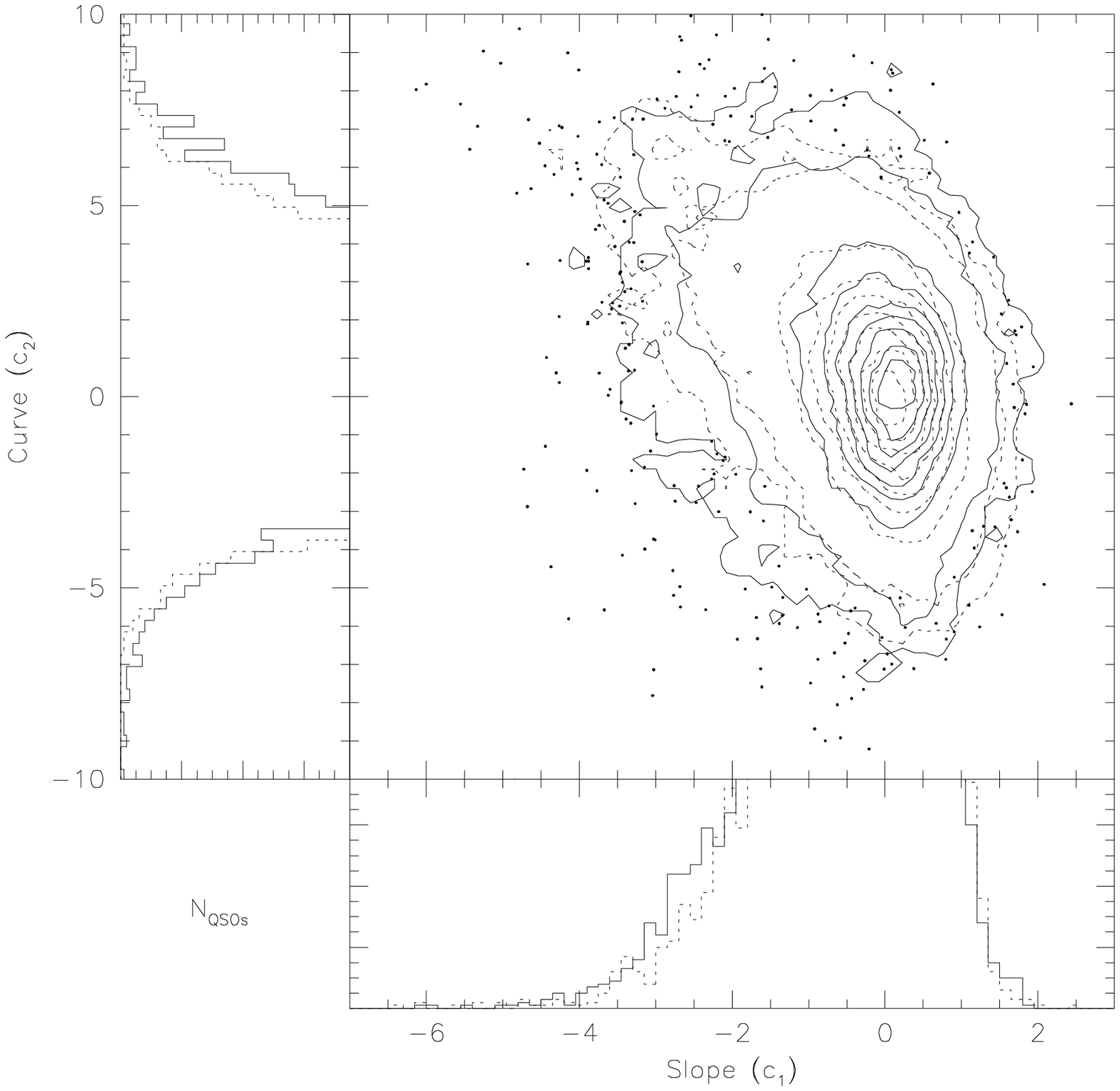}
    \caption{Curvature vs.\ slope ($c_{2}$ vs.\ $c_{1}$) for the 9566
      DR1 quasars, as solid contours. The results of the best-fit
      Monte Carlo simulation are overlaid as dashed contours. The
      histogram below the slope ($c_{1}$) axis shows the marginal distribution
      of observed and simulated quasar slopes, solid and dashed
      respectively. The histogram to the left of the curvature
      ($c_{2}$) axis shows the marginal distribution of observed
      and simulated quasar curvatures.  Note the excellent agreement
      between the observed and model distributions, both in the core
      (as shown by the contours) and the tails (as shown by the
      histograms).\label{simdr1}} 
\end{figure}
\clearpage
\begin{figure}
    \plotone{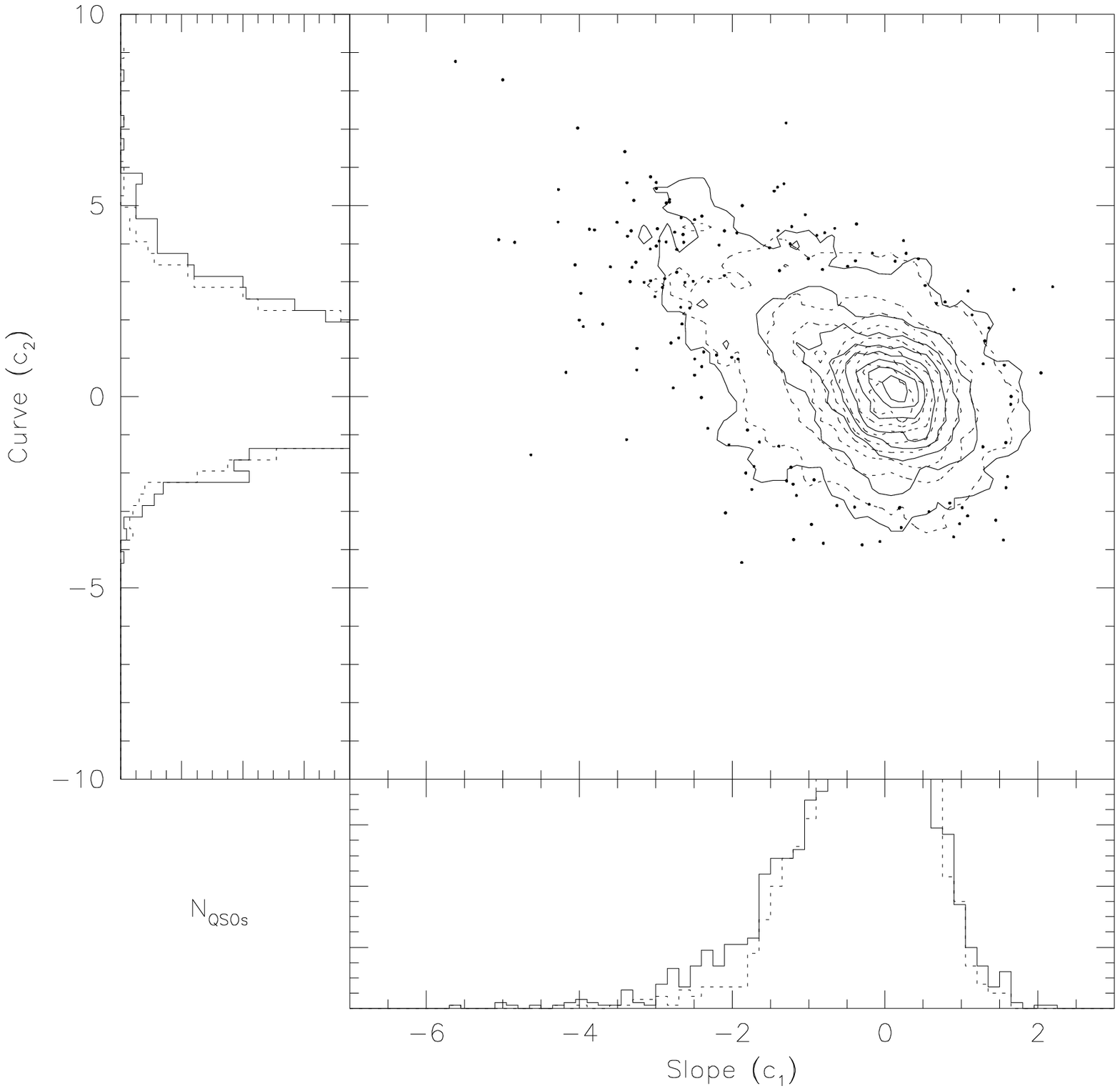}
    \caption{Curvature vs.\ slope ($c_{2}$ vs.\ $c_{1}$) for the 1886
      SDSS-2MASS matched quasars, as solid contours. The results of
      the best-fit Monte Carlo simulation are overlaid as dashed
      contours. The histogram below the slope ($c_{1}$) axis shows the
      marginal distribution of observed and simulated quasar slopes, solid
      and dashed respectively. The histogram to the left of the curvature
      ($c_{2}$) axis shows the marginal distribution of observed
      and simulated quasar curvatures.\label{sim2ma}}
\end{figure}
\clearpage
\begin{figure}
    \plotone{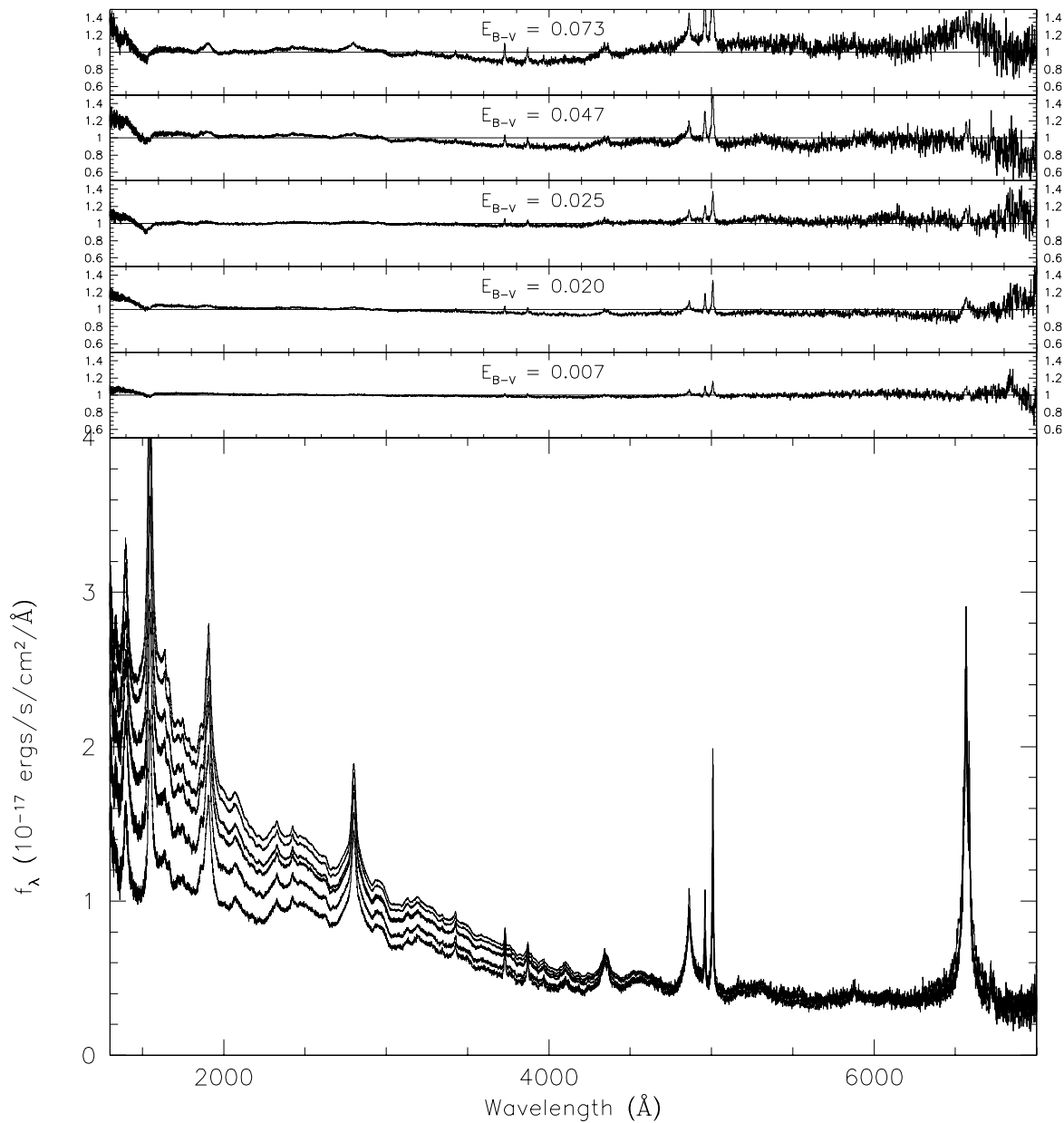}
    \caption{Composite spectra as a function of distance along the SMC
      dust reddening axis $d_{SMC}$, as binned as shown in
      Table~3.  The larger the value of $r_{SMC}$, the more suppressed
      the blue flux is.        Spectra are normalized to constant flux density at 6000\,\AA{}.
      The top half of the plot shows the ratio of each composite to 
      the ``modal'' ($-1<d_{SMC}<1$) composite, reddened with
      an SMC-like dust reddening law with the printed $E_{B-V}$ value, 
      $\frac{f_{\rm composite}}{f_{\rm modal, reddened}}$.
      \label{specfig}}
\end{figure}
\clearpage
\end{center}

\clearpage
\begin{deluxetable}{crrrrrrr}
\tabletypesize{\scriptsize}
\tablecaption{Modal Colors as a Function of Redshift \tablenotemark{a} \label{modestbl}}
\tablewidth{0pt}
\tablehead{
\colhead{$z$} & \colhead{$u-g$} & \colhead{$g-r$} & \colhead{$r-i$} & 
\colhead{$i-z$} & \colhead{$z-J$} & \colhead{$J-H$} & \colhead{$H-K$}}
\startdata
0.10 & -0.026 & 0.279 & 0.549 & -0.128 & --- & 0.824 & 1.009 \\ 
0.15 & -0.028 & 0.166 & 0.427 & -0.071 & --- & --- & --- \\ 
0.20 & -0.004 & 0.314 & 0.382 & -0.006 & 1.329 & 0.761 & 1.042 \\ 
0.25 & 0.048 & 0.305 & 0.213 & 0.196 & --- & --- & --- \\ 
0.30 & 0.016 & 0.190 & 0.047 & 0.449 & --- & 0.799 & 0.996 \\ 
0.35 & 0.039 & 0.133 & 0.004 & 0.462 & --- & --- & --- \\ 
0.40 & 0.185 & 0.022 & 0.084 & 0.364 & 1.002 & 0.764 & 0.939 \\ 
0.45 & 0.244 & -0.102 & 0.163 & 0.158 & --- & --- & --- \\ 
0.50 & 0.291 & -0.069 & 0.177 & 0.056 & --- & 0.724 & 0.803 \\ 
0.55 & 0.279 & -0.126 & 0.139 & -0.025 & --- & --- & --- \\ 
0.60 & 0.307 & -0.029 & 0.160 & -0.073 & 1.112 & 0.607 & 0.822 \\ 
0.65 & 0.302 & -0.030 & 0.113 & -0.068 & --- & --- & --- \\ 
0.70 & 0.331 & 0.009 & 0.029 & 0.054 & --- & 0.516 & 0.819 \\ 
0.75 & 0.289 & 0.026 & -0.021 & 0.105 & --- & --- & --- \\ 
0.80 & 0.223 & 0.035 & -0.102 & 0.098 & 1.118 & 0.260 & 0.750 \\ 
0.85 & 0.208 & 0.062 & -0.073 & 0.110 & --- & --- & --- \\ 
0.90 & 0.172 & 0.135 & -0.068 & 0.107 & --- & 0.303 & 0.790 \\ 
0.95 & 0.134 & 0.153 & -0.053 & 0.072 & --- & --- & --- \\ 
1.00 & 0.124 & 0.196 & -0.047 & 0.081 & 1.165 & 0.305 & 0.786 \\ 
1.05 & 0.119 & 0.221 & -0.041 & -0.016 & --- & --- & --- \\ 
1.10 & 0.098 & 0.224 & -0.032 & -0.071 & --- & 0.558 & 0.624 \\ 
1.15 & 0.043 & 0.226 & 0.014 & -0.054 & --- & --- & --- \\ 
1.20 & 0.041 & 0.246 & 0.023 & -0.084 & 0.879 & 0.553 & 0.567 \\ 
1.25 & 0.044 & 0.242 & 0.000 & -0.080 & --- & --- & --- \\ 
1.30 & 0.009 & 0.241 & 0.030 & -0.054 & --- & 0.714 & 0.323 \\ 
1.35 & 0.072 & 0.221 & 0.028 & -0.039 & --- & --- & --- \\ 
1.40 & 0.027 & 0.218 & 0.069 & -0.036 & 0.916 & 0.750 & 0.209 \\ 
1.45 & 0.071 & 0.116 & 0.118 & -0.039 & --- & --- & --- \\ 
1.50 & 0.157 & 0.113 & 0.177 & -0.039 & --- & 0.739 & 0.216 \\ 
1.55 & 0.200 & 0.093 & 0.203 & -0.034 & --- & --- & --- \\ 
1.60 & 0.176 & 0.064 & 0.215 & -0.018 & 0.940 & 0.733 & 0.180 \\ 
1.65 & 0.156 & 0.040 & 0.225 & -0.016 & --- & --- & --- \\ 
1.70 & 0.119 & 0.018 & 0.226 & 0.015 & --- & 0.561 & 0.419 \\ 
1.75 & 0.091 & 0.013 & 0.246 & 0.017 & --- & --- & --- \\ 
1.80 & 0.020 & -0.010 & 0.257 & 0.006 & 0.929 & 0.479 & 0.586 \\ 
1.85 & 0.031 & 0.031 & 0.246 & 0.031 & --- & --- & --- \\ 
1.90 & 0.044 & 0.030 & 0.211 & 0.053 & --- & 0.598 & 0.555 \\ 
1.95 & 0.005 & 0.034 & 0.174 & 0.103 & --- & --- & --- \\ 
2.00 & 0.017 & 0.048 & 0.133 & 0.145 & 0.792 & 0.482 & 0.121 \\ 
2.05 & 0.133 & 0.084 & 0.117 & 0.183 & --- & --- & --- \\ 
2.10 & 0.085 & 0.069 & 0.109 & 0.189 & --- & 0.613 & 0.723 \\ 
2.15 & 0.258 & 0.113 & 0.079 & 0.202 & --- & --- & --- \\ 
2.20 & 0.390 & 0.057 & 0.071 & 0.209 & 0.811 & 0.601 & 0.748 \\ 
\enddata
\tablenotetext{a}{Modes for other color combinations should be calculated directly but can
be estimated within 7\% rms from the modes given in this Table.}
\end{deluxetable}

\clearpage
\begin{deluxetable}{clrcccccc}
\tabletypesize{\scriptsize}
\tablecaption{Fit parameters to Slope and Curvature Distribution \label{simtbl}}
\tablewidth{0pt}
\tablehead{
\colhead{Sample} & \colhead{Profile} & \colhead{N$_{QSOs}$} & \colhead{$\sigma_{\alpha}$} & 
\colhead{$\sigma_{error}$} & \colhead{$\sigma_{dust}$} & \colhead{$\sigrat$} & 
\colhead{$n$} & \colhead{D}}
\startdata
SDSS DR1 & $\exp(-x)$ & 9566 & 0.11 $\pm$ 0.010 & 0.065 $\pm$ 0.010 & 0.032 $\pm$ 0.005 & 0.54 $\pm$ 0.03 & --- & 0.047 \\
$\phn$ & $(1+x^{n})^{-1}$ & 9566 & 0.13 $\pm$ 0.015 & 0.070 $\pm$ 0.020 & 0.027 $\pm$ 0.005 & 0.55 $\pm$ 0.05 & 2.5 $\pm$ 0.02 & 0.059 \\
SDSS-2MASS & $\exp(-x)$ & 1886 & 0.13 $\pm$ 0.015 & 0.180 $\pm$ 0.020 & 0.045 $\pm$ 0.010 & 0.55 $\pm$ 0.05 & --- & 0.043 \\
$\phn$ & $(1+x^{n})^{-1}$ & 1886 & 0.13 $\pm$ 0.020 & 0.185 $\pm$ 0.010 & 0.042 $\pm$ 0.007 & 0.60 $\pm$ 0.10 & 2.3 $\pm$ 0.02 & 0.044 \\
\enddata
\end{deluxetable}

\clearpage

\begin{deluxetable}{rrrr}
\tabletypesize{\scriptsize}
\tablecaption{Composite Spectra Bins \label{spectbl}}
\tablewidth{0pt}
\tablehead{
\colhead{Minimum} & \colhead{Maximum} & 
\colhead{Number} & \colhead{Bestfit}\\
\colhead{$d_{SMC}$} & \colhead{$d_{SMC}$} & 
\colhead{of QSOs} & \colhead{$E_{B-V}$}
}
\startdata
$-$1.0 & 1.0 & 9475 & 0.000 \\
1.0 & 3.0 & 6971 & 0.007 \\
3.0 & 5.0 & 4074 & 0.020 \\
5.0 & 7.5 & 2375 & 0.025 \\
7.5 & 10.0 & 1261 & 0.047 \\
10.0 & 15.0 & 1050 & 0.073 \\
\enddata
\end{deluxetable}

\end{document}